\documentclass[twocolumn,showpacs,preprintnumbers,amsmath,amssymb]{revtex4}
\usepackage{graphicx}
\usepackage{dcolumn}
\usepackage{bm}

\begin{document}

\title{Crossover from Thermal Activation to Quantum Interlayer Transport at the
Superconducting Transition Temperature of Bi$_2$Sr$_2$CaCu$_2$O$_{8+\delta}$}

\author{S. O. Katterwe}
\author{A. Rydh}
\author{V. M. Krasnov}
\email{Vladimir.Krasnov@physto.se}

\affiliation{Department of Physics, Stockholm University, AlbaNova
University Center, SE-10691 Stockholm, Sweden}

\date{\today }

\begin{abstract}

We perform a detailed study of temperature, bias and doping
dependance of interlayer transport in the layered high
temperature superconductor Bi$_2$Sr$_2$CaCu$_2$O$_{8+\delta}$. We
observe that the shape of interlayer characteristics in underdoped crystals exhibit a
remarkable crossover at the superconducting transition temperature: from thermal activation-type at $T>T_c$, to almost $T-$independent quantum tunneling-type, at $T<T_c$. Our data indicates that the interlayer transport mechanism may change with doping: from the conventional single quasiparticle tunneling in overdoped, to a progressively increasing Cooper pair contribution in underdoped crystals.


\end{abstract}

\maketitle

How does the high temperature superconductivity (HTSC) emerge
with decreasing temperature? This highly debated question is
crucial for understanding the mechanism of HTSC. The
superconducting transition in HTSC appears to be unusual, with a
seeming lack of changes in the quasiparticle (QP) density of
states at $T_c$ and persistence of the normal state
pseudogap above $T_c$ \cite{TallonPhC,Timusk}.
However, there is no consensus about the $T-$evolution of the energy gap. Results obtained by
different experimental techniques range from complete $T-$
independence of the superconducting gap, $\Delta_{SG}$, in the
whole $T-$ range \cite{Renner}, to strong $T-$dependence at
$T=T_c$ \cite{BreakJun,Ponomarev,PointCont,KrTemp,Doping,Lee2007}. In some experiments,
coexistence of the two energy scales below $T_c$ was reported \cite{KrTemp,Doping,Lee2007,Raman}. The existing controversy is one of the major obstacles for understanding HTSC
and requires further $T-$dependent studies.

Intrinsic tunneling spectroscopy utilizes the weak
interlayer ($c-$axis) coupling in layered HTSC. Unlike surface
probe techniques, it is perfectly suited for $T-$dependent
studies and provides information about bulk electronic properties
of HTSC. In recent years different ways of improving this technique were
employed \cite{KrTemp,Doping,Suzuki,Latyshev,Lee,Cascade}.
Yet, interpretation of intrinsic tunneling characteristics may not be free from
misconceptions, associated both with the problem of self-heating
\cite{Suppl} and the lack of complete understanding of the
interlayer transport mechanism.

Here we study temperature, bias and doping dependencies of intrinsic tunneling
in small Bi$_2$Sr$_2$CaCu$_2$O$_{8+\delta}$
(Bi-2212) mesa structures. We observe an abrupt crossover in the shape of current-voltage characteristics (IVC's), from thermal activation (TA) like at $T>T_c$, to $T-$independent quantum tunneling like at $T<T_c$. Our data indicates that the interlayer transport mechanism in Bi-2212 changes in underdoped Bi-2212, together with development of the pseudogap. The observed simple TA behavior in the whole normal state region puts strong constrains on the nature of the $c-$axis pseudogap. In the supplementary information \cite{Suppl} we  discuss single QP characteristics, self-heating and clarify determination of parameters.

We study two batches of crystals: the Y-doped
Bi(Y)-2212 with the maximum $T_c\simeq 94.5$K and the pure Bi-2212
with the maximum $T_c \simeq 86$K. Details of mesa fabrication and characterization are described elsewhere \cite{KrTemp,Doping,Cascade}.

Figs. 1 a) and b) show the $T-$ variation of $I-V$ and $dI/dV(V)$ curves for a Bi-2212 mesa, containing $N=7$ intrinsic Josephson junctions, see Fig. 1 c). The following characteristic features are seen in $dI/dV$ curves: The sharp "coherence" peak appears at $T<T_c$ and is, in analogy with
conventional low-$T_c$ junctions, attributed to the sum-gap voltage $V_{g}=2N\Delta_{SG}/e$
\cite{KrTemp,Doping,Suzuki,Lee}. A broader hump is seen in the whole $T-$ range and is attributed to the $c-$ axis pseudogap \cite{KrTemp,Doping,Suzuki}. At elevated $T$ the $dI/dV$ curves cross in nearly one point \cite{Doping,Lee}. The $T-$dependence of these features is plotted in Fig. 1 d)

Figure 2 a) shows $dI/dV(V)$ curves (in a semi-log scale) for the most underdoped Bi(Y)-2212 mesa with $N=34$ junctions. The thin and thick lines in Fig. 2 a) represent $dI/dV(V)$ curves below and
above $T_c$, respectively. It is seen that the shape of interlayer characteristics exhibits a remarkable qualitative change at $T_c$:

(i) Above $T_c$ the slope of $\ln [dI/dV](V)$ curves decreases with increasing
$T$, while the voltage scale remains almost the same. The curves lean
towards the horizontal line $dI/dV = \sigma_N$, corresponding to the normal state conductance, and cross almost in one point.

(ii) Below $T_c$ the $\ln [dI/dV](V)$ curves have {\it the same} $T-${\it independent slope}. With increasing $T$ the curves remain parallel and move towards the vertical axis $V=0$ (i.e., the voltage scale decreases).

\begin{figure}
\includegraphics[width=3.5in]{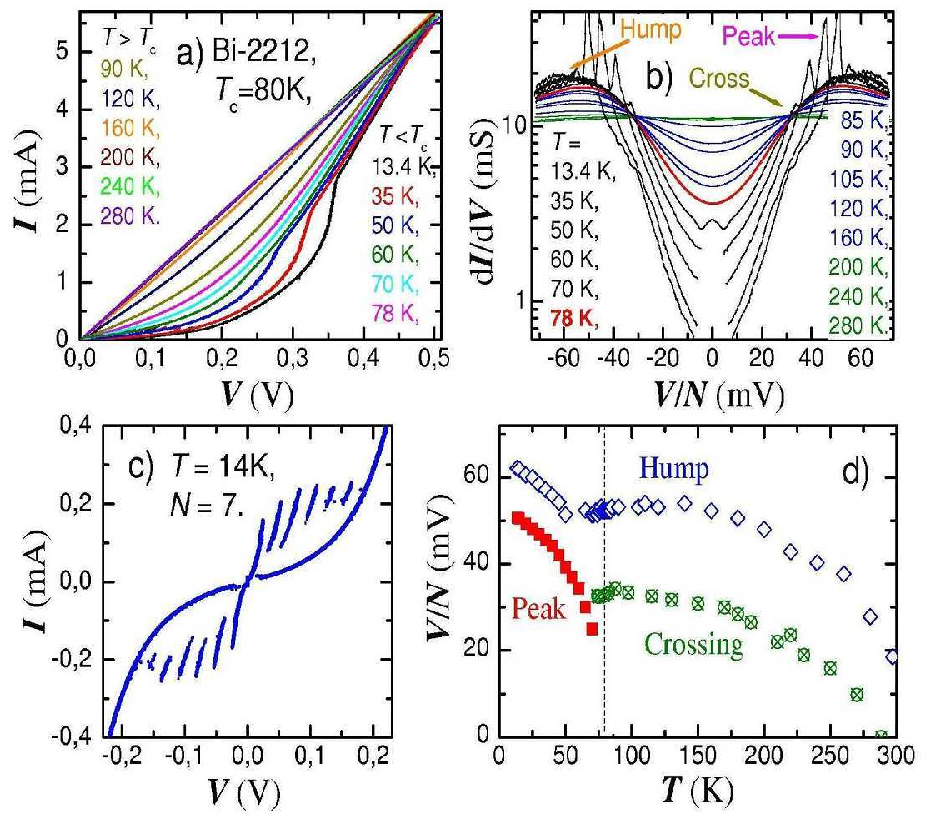}
\caption{(Color online). a) $I-V$ and b) $dI/dV(V)$ characteristics for a
Bi-2212 mesa at different $T$. c) Periodic quasiparticle branches.
d) Temperature dependence of the peak, hump and crossing voltages (per junction).}
\end{figure}

Experimental curves from Fig. 2 a) have a characteristic V-shape with voltage-independent slope. The overall shape of the curves at $T>T_c$ closely resembles thermal activation characteristics  \cite{Silva}, which can be described by a simple expression \cite{Suppl}:
\begin{equation}
\frac{dI}{dV}(T,V) \propto \frac{n(T)}{T}
\exp\left[-\frac{U_{TA}}{k_B T}\right]\cosh\left[\frac{eV}{2k_B T}\right], \label{TeffN}
\end{equation}
where $n$ is the concentration of mobile charge carriers, and $U_{TA}$ is the TA barrier. Indeed, the $cosh$ term reproduces the rounded V-shape of $\ln [dI/dV](V)$ curves with the slope that monotonously increases as $1/T$.

In Fig. S10 \cite{Suppl} we show that the data from Fig. 2 a) in the whole normal state, $T>T_c$, is very well described by Eq.(\ref{TeffN}) with constant $U_{TA}\simeq 24 meV$ and $T-$linear $n(T)$. It also reproduces the crossing point \cite{Silva}, which according to Eq.(\ref{TeffN}) occurs at $eV \simeq 2U_{TA}$. Apparently, the TA barrier should be associated with the $c-$axis pseudogap. Thus, the appearance of the crossing point is simply the consequence of a fairly $T-$independent $c-$axis pseudogap \cite{TallonPhC,Renner,KrTemp,Doping}, while the crossing voltage represents the pseudogap energy. Indeed, a correlation between the hump and the crossing voltages is seen from Fig. 1 d).

According to Eq.(\ref{TeffN}) the slope of the curves in Fig. 2 a) should be proportional to the reciprocal temperature: $d/dV(\ln [dI/dV])=$ $(e/2k_BT) \tanh (eV/2k_BT)$ $\simeq e/2k_BT$.
Fig. 2 b) shows the effective temperature, $T_{eff}$, obtained
from the slopes at $V/N=$30mV \cite{Suppl}. It is seen that in the normal state $T_{eff} \simeq T$, confirming the TA nature of interlayer transport at $T>T_c$. However, at $T<T_c$ an abrupt saturation of $T_{eff}(T)$ occurs, typical for quantum tunneling transport \cite{Clark,Collapse}. This is the central observation of this work. Remarkably, the crossover is clearly distinguishable at $T\rightarrow T_c$, although no other spectroscopic features can be resolved in $dI/dV(V)$ curves.


\begin{figure}
\includegraphics[width=2.6in]{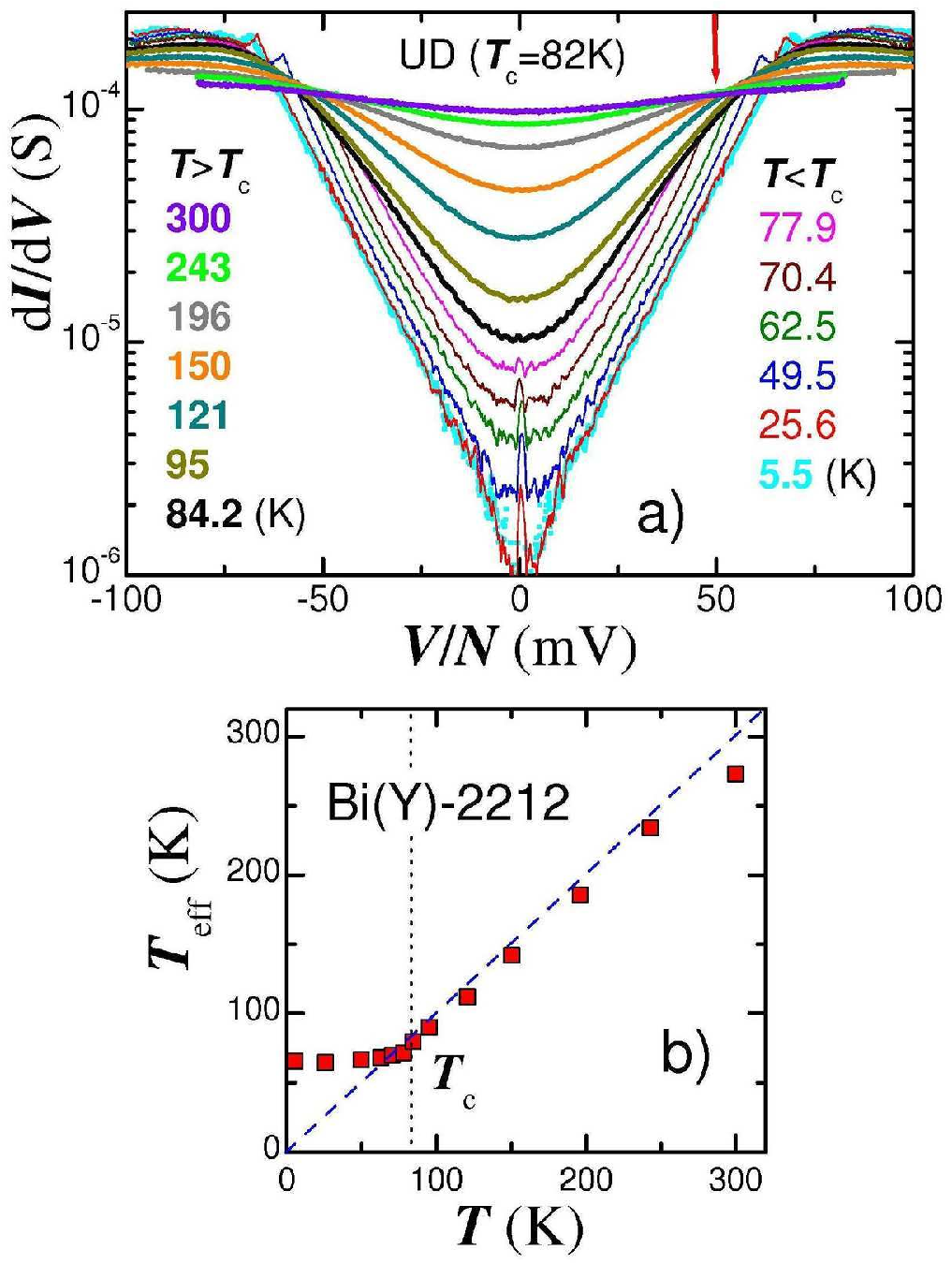}
\caption{(color online). a) A semi-log plot of $dI/dV(V)$ curves
for an underdoped Bi(Y)-2212 mesa at different $T$. Note that below $T_c$ the curves maintain the same slope, while above $T_c$ the slope changes progressively with $T$. A characteristic crossing
point of $dI/dV(V)$ curves at $T>T_c$ is marked by the arrow.
b) The effective temperature, obtained from the slopes of the
$dI/dV(V)$ curves at a finite bias $V/N=30 mV$ \cite{Suppl}. A
remarkable crossover from the thermal activation, $T_{eff}=T$, to
quantum, $T_{eff}=\textrm{const}$, behavior occurs at $T_c$.
}
\end{figure}

So far we discussed $dI/dV(V)$ at finite bias. Another straightforward way to investigate the TA behavior is to consider $T-$dependence of zero-bias resistance, $R_0$, (the $exp$ term in Eq.(\ref{TeffN})). Experiments on small mesas provide a possibility to measure the small-bias QP resistance, $R^{QP}$, in the superconducting state \cite{Suzuki,YurgensPRL,Latyshev}, which is otherwise shunted by the supercurrent \cite{Morozov}. This is facilitated by the large specific capacitance (high quality factor) of intrinsic junctions \cite{Collapse}, which causes the pronounced hysteresis in the IVC's and allows junctions to remain in the resistive state even below the critical current, see Fig. 1 c). Therefore, bias yield an additional parameter for our studies, which may render crucial for correct interpretation of the data.

Fig. 3 a) shows the $dc-$ resistance $R_{dc}=V/I$ at different $I$ for the same Bi-2212 meas as in Fig. 1. Measurements were done by first applying a large current, sufficient for switching to the last QP branch, see Fig. 1 c), and then ramping it back to the desired value. The $R_{dc}$ at the smallest current $I=5 \mu A$ coincides with the conventional $ac-$ resistance, $R_{ac}$. It drops at $T<T_c$, since the finite quality factor of the junctions causes retrapping to the superconducting branch at a finite current \cite{Collapse}. From Fig. 3a) it is seen how non-linearity develops with decreasing both bias and $T$. At small $I$ the curves approach an asymptotic, allowing a confident estimate of the zero-bias quasiparticle resistance, $R_0^{QP}$, without a need for extrapolation.

\begin{figure}
\includegraphics[width=3.5in]{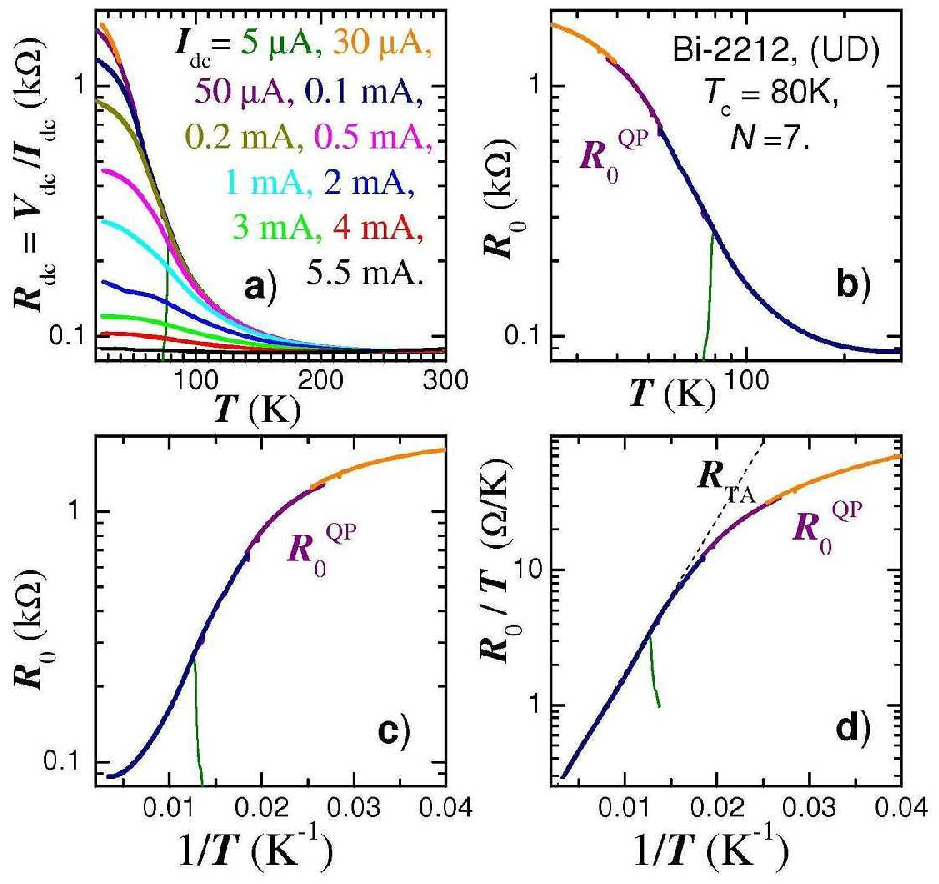}
\caption{(Color online). a) $T-$ dependence of the $dc-$ resistance at different bias currents for the same Bi-2212 mesa. Panels b-d) show the asymptotic zero bias resistance: b) in the double-logarithmic scale; c) as a function of $1/T$; and d) normalized by $T$: a clear linear thermal activation behavior (dashed line) is observed in the whole normal state region.}
\end{figure}

The $R_0^{QP}$ grows with decreasing $T$ at $T<T_c$, and is usually described either in terms of power-law dependence \cite{Morozov}, inherent for single QP tunneling in the presence of a $d-$wave gap (see Fig. S7 \cite{Suppl}), or in terms of TA \cite{Watanabe}. The double-logarithmic plot, Fig. 3 b), demonstrates that $R_{0}(T)$ is not described by the power law in any extended $T-$range for our mesas. Neither is it perfectly described by the Arrhenius law $exp(U_{TA}/k_B T)$, as demonstrated in Fig. 3 c). On the other hand, Fig. 3 d) demonstrates that the ratio $R_0/T$ follows very accurately the Arrhenius law (dashed line) in the whole normal region, consistent with the TA expression Eq.(\ref{TeffN}) and with the finite bias behavior, Fig. 2 b). However, at $T<T_c$, $R_0^{QP}/T$ deviates downwards from the Arrhenius law. This is consistent with saturation of the effective $T_{eff} (T<T_c)$, as shown in Fig. 2 b). Therefore, it is a consequence of the same crossover.

\begin{figure}
\includegraphics[width=3.2in]{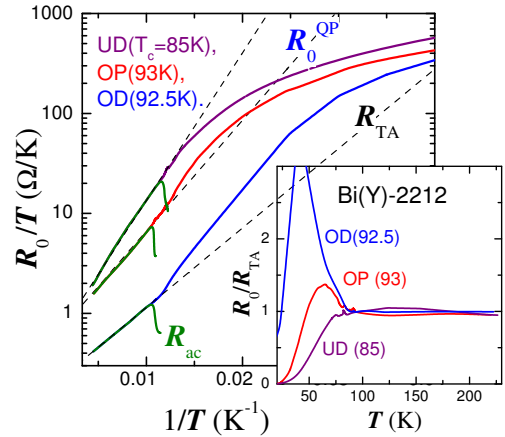}
\caption{\label{Fig1} (Color online). $T-$normalized zero-bias quasiparticle
resistance, $R_0^{QP}$, extrapolated from IVC's at $T<T_c$ and $R_{ac}$ measured with a small
ac-current for Bi(Y)-2212 mesas with different doping. Dashed lines represent TA fits, Eq.(\ref{TeffN}), at $T>T_c$. Inset shows relative values of the excess QP resistance with respect to the TA fit of the normal state. A large excess resistance appears below $T_c$ in the overdoped mesa. The excess resistance rapidly decreases with decreasing doping and becomes negative in the underdoped mesa.}
\end{figure}

In Fig. 4 we analyze doping dependence of $R_{0}(T)/T$. In the normal state, $R_0(T)$ varies in the TA manner for all doping levels. In the superconducting state, $R_0^{QP}$ also continues to grow with decreasing $T$, but the rate of the growth with respect to the TA behavior, $R_{TA}$, (dashed lines) strongly depends on doping:

For the overdoped (OD) mesa, $R_0^{QP}(T<T_c)$ grows much faster than in the normal state, i.e., there is a sharp onset of the excess resistance at $T=T_c$. However, the excess resistance at $T<T_c$ decreases rapidly with underdoping. It becomes small in the near optimally doped (OP) mesa and turns {\it negative} already in moderately underdoped (UD) mesas.

Inset in  Fig. 4 shows the magnitude of the excess QP resistance with respect to the TA fit, $R_{TA}$, in the normal state (dashed lines). The quality of the fit could be judged from the flatness of the normal state region.
A progressive decrease of the excess QP resistance at $T<T_c$ with decreasing doping is obvious.

We start discussion of the observed phenomena by recollecting the expected $T-$ variation for conventional single QP tunneling characteristics. As discussed in sec. I of the supplementary information, provided the physical requirement that $\Delta_{SG}$ and the quasiparticle lifetime increase with decreasing $T$, all single QP tunneling characteristics (including $d-$wave) exhibit the following universal $T-$dependence:

A) Opening of the superconducting gap at $T<T_c$ leads to appearance of the large excess QP resistance due to rapid freezing out of QP's upon their condensation into Cooper pairs (see inset in Fig. S7 \cite{Suppl}).

B) $R_0^{QP}$ continues to grow with decreasing $T$ both due to freezing-out of thermal QP's and due to growth of the QP lifetime, which decreases the number of available sub-gap QP states. All this leads to a progressive growth of the slope of $dI/dV(V)$ characteristics with decreasing $T$.

The sudden appearance of excess resistance at $T<T_c$ in OD mesas indicates that the superconducting transition here is conventional, in a sense that it is accompanied by opening of the superconducting gap, and is also consistent with single QP tunneling mechanism of interlayer transport.

However, the superconducting transition in underdoped mesas is clearly abnormal. We emphasize again that appearance of the negative excess resistance and the crossover to $T-$independent slope reflect the same phenomenon at zero and finite bias, respectively. Therefore, any explanation of the observed unusual $T-$dependence must account for both zero and finite bias behavior.

At the first glance, the lack of excess resistance at $T<T_c$ in UD mesas could be consistent with the precursor superconductivity scenario of the pseudogap \cite{Renner}, according to which the gap does not open and QP's do not start to pair at $T_c$, but at much higher $T^*$. Then one could argue that the lack of excess resistance reflects simply the lack of dramatic changes in the QP spectrum upon establishing of the global phase coherence at $T_c$. However, the same argument would make it difficult to explain the abrupt crossover to $T-$independent slope. Furthermore, as we noted above, the slope of $dI/dV(V)$ should continue to increase with decreasing $T$ for single QP tunneling. Therefore, the observed crossover to constant slope can hardly be explained in terms of single QP tunneling \cite{Suppl}.

This conclusion brings us to one possible interpretation of the observed phenomena. If the $c-$axis transport in UD mesas is not solely due to single QP's, it must also involve pairs. The corresponding multi-particle tunneling \cite{Taylor,Rowell} and the multiple Andreev reflection \cite{Kleinsasser,Bratus} processes are well studied for conventional low$-T_c$ junctions.
Both processes are similar \cite{Bratus} and are almost
$T-$ independent, because they do not rely on the presence of
thermally excited QP's, but directly involve Cooper pairs from the Fermi level.
The multi-particle current occurs via elastic conversion (dissociation or
recombination) of Cooper pairs into QP's;
while the multiple Andreev reflection process is due to transmutation of a quasielectron
into a quasihole with creation of a Cooper pair and can be
universally described in terms of inelastic tunneling in the
presence of time-dependent phase difference due to the
ac-Josephson effect \cite{Bratus}.

Importantly, the multi-particle processes result in almost $T-$independent and $V-$exponential
subgap current \cite{Taylor,Rowell}, which essentially coincides with the description of the observed crossover, see Fig. 2. Furthermore, the abruptness of the crossover at $T_c$, points towards the coherent multiple Andreev reflection mechanism of the interlayer current, because it requires the time-periodic ac-Josephson effect \cite{Bratus} and,
consequently, abruptly disappears simultaneously with phase
coherence at $T_c$.

We may further speculate why the multiparticle processes become progressively more important with underdoping. The multiparticle current decrease much faster than the single QP current
with decreasing the interface transparency \cite{Taylor,Bratus}.
Therefore, the single QP dominates over multiparticle current, unless there are micro-shots in the tunnel barrier \cite{Rowell}. The required micro-shots in underdoped Bi-2212 could arise from
doping-dependent nano-scale inhomogeneity, observed at the surface
of Bi-2212 \cite{Davis,ImpSTM}, which in this case should persists also in
the bulk of Bi-2212 crystals.

Such interpretation implies that not only the
electronic structure, but also the $c-$axis {\it transport
mechanism} changes with underdoping: from coherent and directional \cite{Suppl} {\it single} QP tunneling in overdoped, to progressively increasing {\it pair} contribution in underdoped Bi-2212. The latter is apparent
only in the phase coherent state at $T<T_c$, is almost
$T-$independent, and is consistent with multiple
Andreev reflection mechanism of the interlayer transport.

Finally we note that the reported amazingly trivial thermal activation behavior in the whole normal state region, puts strong constrains on the nature of the $c-$axis pseudogap. Even though, the similar TA-behavior can be ascribed to several processes (e.g., inelastic tunneling via the impurity \cite{Impurity,ImpSTM}, or elastic tunneling via a resonant \cite{Abrikosov} state in the tunnel barrier, or Coulomb blocking of tunneling \cite{KrTemp}), it does not involve any gap, nor angular dependence in the QP spectrum, but assumes instead that there is some constant blocking barrier for interlayer hoping \cite{Silva}. This may indicate that the "large" $c-$axis pseudogap is not a pairing gap, which would naturally explain it's indifference to such classical depairing factors as temperature and magnetic field \cite{KrTemp}.

We are grateful to E.Silva for stimulating discussions and to A.Yurgens, T.Benseman and G.Balakrishnan for providing Bi(Y)-2212 and Bi-2212 single crystals. Financial support from the
K.{\&}A. Wallenberg foundation and the Swedish Research Council is gratefully acknowledged.

\section{SUPPLEMENTARY INFORMATION}

\subsection{Nomenclature}

ARPES: angular resolved photoemission spectroscopy,

Bi-2212: Bi$_2$Sr$_2$CaCu$_2$O$_{8+\delta}$,

Bi(Y)-2212: Y-doped Bi-2212,

DoS: density of states,

$\Delta$: energy gap,

$\Delta T$: self-heating,

$\Gamma$: depairing factor, reciprocal quasiparticle lifetime;

HTSC: high temperature superconductor,

$I$: current,

IJJ: intrinsic Josephson junction.

IVC: Current-Voltage characteristic,

$N$: number of IJJ's in the mesa,

$n$: concentration of mobile charge carriers,

OD, OP, UD: overdoped, optimally doped, underdoped, respectively;

QP: quasiparticle,

$R_0$: zero bias resistance (in the normal state),

$R_0^{QP}$: zero bias quasiparticle resistance in the superconducting state, is different from $R_0$ because of the pronounced hysteresis in the IVC's at $T<T_c$;

$R_n$: tunnel (normal) resistance,

$R_{th}$: thermal resistance of a mesa,

SIS: Superconductor-Insulator-Superconductor,

SIN: Superconductor-Insulator-Normal metal,

STM: scanning tunneling microscope,

TA: thermal activation,

$T_c$: superconducting critical temperature,

$U_{TA}$: thermal activation barrier,

$V$: voltage,

$V_g$: the sum-gap voltage $V_g=2N\Delta/e$

The capital M/S in front of the figure or equation number indicates that the reference is made to the Manuscript or the Supplementary information, respectively.

\subsection{Influence of the quasiparticle life time on $dI/dV$ characteristics}

We have argued in the manuscript that the observed crossover to $T-$independent slope and the correlated appearance of the negative excess resistance are inconsistent with the single QP tunneling mechanism of the $c-$axis transport. However, the interlayer tunneling in Bi-2212 may depend on a number of parameters \cite{Millis,Carbotte,Yamada}:

(i) The QP life time, $1/\Gamma$, which, according to ARPES data, has a substantial $T-$dependence close to $T_c$, where it increases roughly linearly with $T$; \cite{Norman,Lee2007}

(ii) Momentum conservation upon tunneling (coherent vs. incoherent tunneling);

(iii) The directionality of $c-$axis tunneling, i.e., the angular dependence of the tunneling matrix element $t(\varphi_1,\varphi_2)$, caused by a non-spherical Fermi surface \cite{Millis,Carbotte}.

(iv) Temperature and (v) angular dependence of the gap in the DoS and the points (i-iii) above. Deviations from pure $d-$wave symmetry were reported in recent ARPES experiments \cite{Lee2007}.

It may not be obvious that experimental data can not be described by the single QP tunneling with a fortunate combination of all those parameters. Therefore, below we want to dwell upon this statement.

We start with considering the consequences of points (ii, iii) and (v), i.e., coherence and directionality of tunneling and the symmetry of the order parameter. The single QP tunneling current is given by:
\begin{eqnarray}\label{Eq.Tunn}
I(V) = \int_0^{2\pi} \int_0^{2\pi} \int_{-\infty}^{+\infty} d\varphi_1 d\varphi_2 dE~~~~~~~~~ ~ \\
\nonumber
t(\varphi_1,\varphi_2)\rho(E,\varphi_1)\rho(E+eV,\varphi_2)f(E)\left[1-f(E+eV) \right],
\end{eqnarray}

where $E$ is the energy of the QP with respect to the Fermi surface, $\varphi_{1,2}$ are the angles in the momentum space of the initial and the final state of the QP, and $\rho(E,\varphi_1)$ $\rho(E+eV,\varphi_2)$ are the corresponding QP DoS:
\begin{equation}\label{N(G)}
\rho(E,\varphi)=\Re[(E-i\Gamma)/\sqrt{(E-i\Gamma)^2-\Delta(\varphi)^2}],
\end{equation}

In Fig. S5 a) and b) we show numerically simulated SIS characteristics for different single QP tunneling scenarios. We assumed, $t(\varphi_1,\varphi_2)=$const for incoherent-nondirectional, $t(\varphi_1,\varphi_2) \propto \delta(\varphi_1-\varphi_2)$ for coherent-nondirectional and $t(\varphi_1,\varphi_2) \propto [\cos(k_x)-\cos(k_y)]^2 \delta(\varphi_1-\varphi_2)$ for coherent-directional tunneling \cite{Millis}. The simulations were made for $\Delta(\varphi = 0) = 35 meV$, at low $T$ and small depairing $\Gamma \ll \Delta(0)$. Detailed discussion of $dI/dV$ characteristics can be found in Ref. \cite{Yamada}. Irrespective of the scenario, the zero-bias resistance diverges (the conductance tends to zero) at $T\rightarrow 0$ and $\Gamma=0$. The divergence is removed by the finite $\Gamma$, see the inset in Fig. S6 b). Therefore, $dI/dV(V=0,T\rightarrow 0)$ provides the information about the value of $\Gamma$. We also note that $dI/dV(V)$ characteristics exhibit a sharp peak at the sum-gap voltage $V_g=2\Delta/e$ ( except for the curve A).

It is expected that interlayer QP tunneling in Bi-2212 single crystals should be predominantly coherent (provided that the single crystal is pure enough so that there is no momentum scattering upon tunneling) and strongly directional, with dominating antinodal tunneling \cite{Millis,Carbotte}. Indeed, the curve C in Fig. S5 resembles most closely the experimental characteristics (cf. with the curves at $T=13.4K$ from Fig. M1a and b), see also a discussion in Refs. \cite{KrPhysC,Yamada}.

\begin{figure}\label{FigS1}
\includegraphics[width=2.5in]{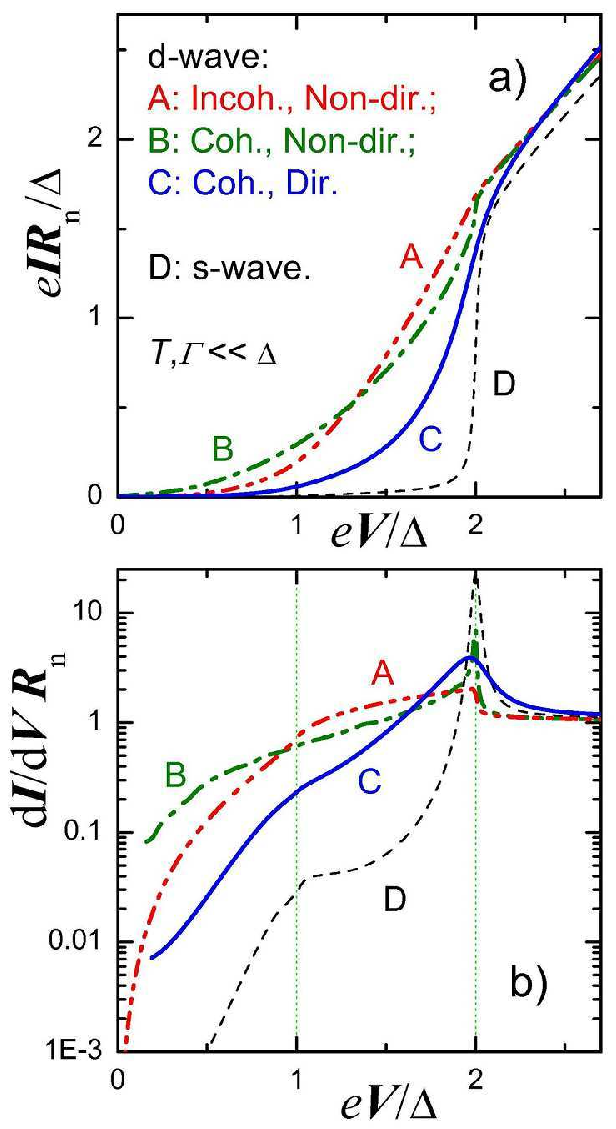}
\caption{(Color online). Simulated single QP characteristics for different tunneling scenarios. A: incoherent, non-directional, $d-$wave; B: coherent, non-directional, $d-$wave; C: coherent, directional, $d-$wave; D: $s-$wave. Note that the sharp sum-gap peak and the half-gap singularity in $dI/dV$ are inherent to most scenarios.}
\end{figure}

In what follows, we will analyze the incoherent-nondirectional single QP tunneling characteristics, representing an "extreme" $d-$wave case with the weakest sum-gap peak. The conclusions will, however, be universal for any single QP tunneling scenario.

Now we consider the effect of $T$ (point iv) and $\Gamma$ (point i) on $dI/dV$ characteristics. It is clear that both parameters will smoothen $dI/dV(V)$ characteristics and fill-in the dip in conductance at $V=0$. However, they do this in a slightly different manner. This is illustrated in Figs. S6 and S7.

\begin{figure}
\includegraphics[width=3.0in]{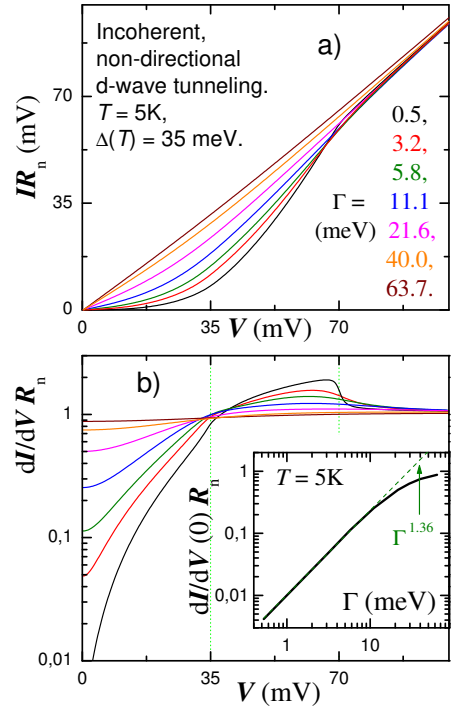}
\caption{(color online). Effect of the depairing factor on single QP characteristics. Shown are simulated characteristics at constant $T$ and $\Delta$ for varying $\Gamma$. Increase of $\Gamma$ tends to reduce the slope of $dI/dV$ characteristics. Inset in panel b) indicates that filling-in of the zero bias dip with $\Gamma$ occurs in a power law manner.}
\end{figure}

The increase of $\Gamma$ leads to smearing of the QP gap singularity and simultaneous filling-in of the sub-gap states in DoS, Eq.(S2). This naturally leads to smearing of the sum-gap peak and to filling-in of the zero-bias dip in $dI/dV$, as shown in the inset in Fig. S6 b).

\begin{figure}
\includegraphics[width=3.0in]{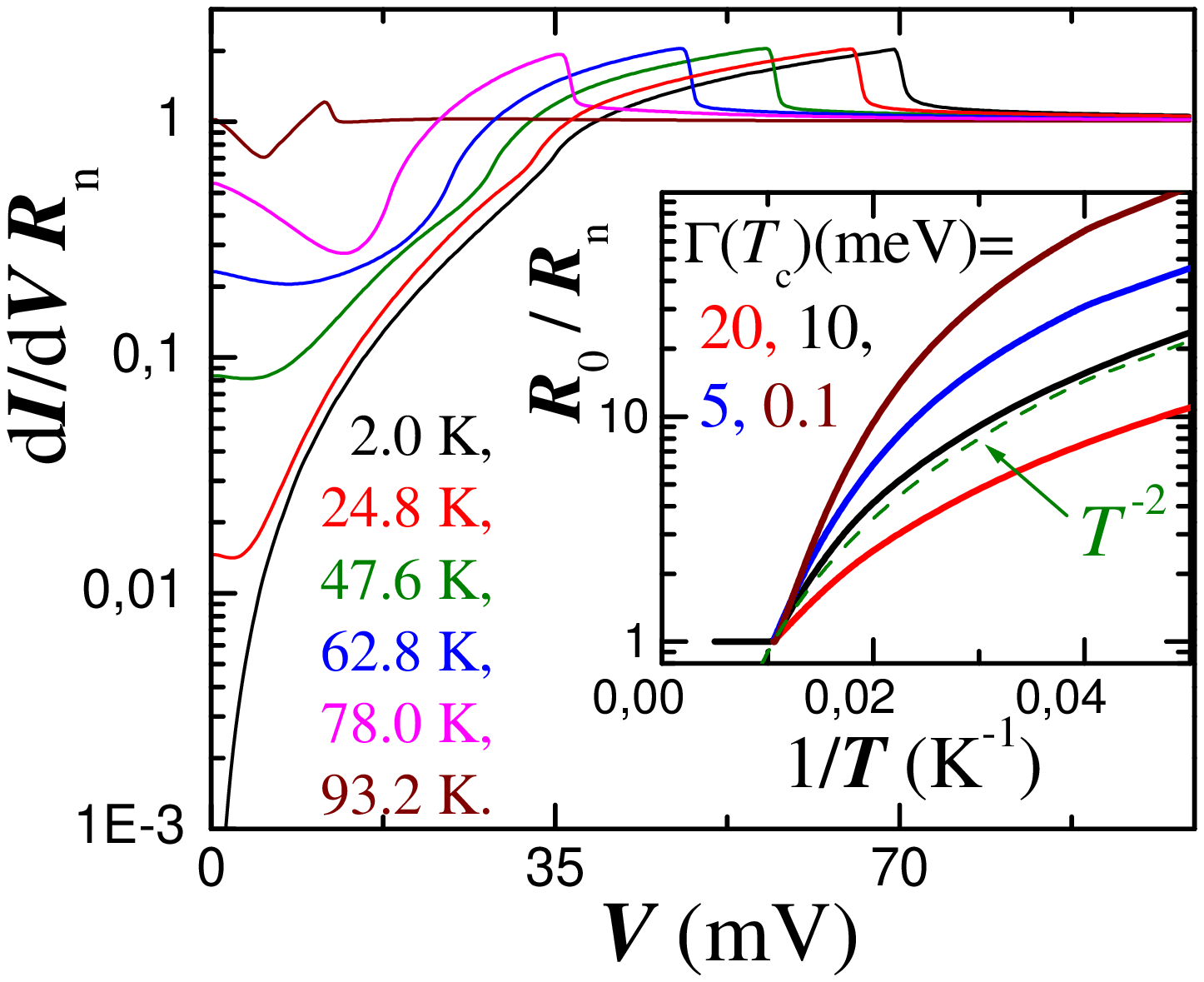}
\caption{(color online). Effect of $T$ and $\Delta(T)$ variation on single QP characteristics. Calculations were made for incoherent, non-directional, $d-$wave tunneling, the experimental $\Delta(T)$ dependence and linear $\Gamma(T)$. Note that increase of $T$ and decrease of $\Delta$ leads not only to smearing of $dI/dV$ characteristics, but also the appearance of the logarithmic zero-bias singularity. Inset shows the zero bias resistance vs. $1/T$ for different values of $\Gamma(T_c)$. It is seen that the excess QP resistance rapidly develops at $T<T_c$ even in the $d-$wave case. }
\end{figure}

The increase of $T$ leads to the decrease of $\Delta$. Both factors result in the increased number of excited QP's above the gap, which leads to rapid filling-in of the zero bias dip in conductance. However, the increase of $T$ (unlike $\Gamma$) keeps the gap singularity in DoS unchanged. From Fig. S7 it is seen that this leads not only to filling-in of the dip, but also to development of the maximum at $V=0$, representing the so called zero-bias logarithmic singularity \cite{Larkin}.

The origin of the singularity is very simple: at elevated $T$ there is a substantial amount of thermally excited QP's just above the gap. At $V=0$ the partly filled gap singularities in the two electrodes are co-aligned, causing a large current flow from one electrode to another, which is exactly compensated by the counterflow from the second electrode. However, exact cancelation is lifted at an arbitrary small voltage across the junction, leading to a sharp maximum in $dI/dV$. From Fig. S7 it is seen that the zero-bias maximum in $d-$wave junctions is pronounced even in the extreme case of weakest sum-gap peak. We have checked that the zero-bias logarithmic singularity is inherent for single QP characteristics, irrespective of the tunneling scenario.

Interestingly, the experimental characteristics do not exhibit the zero-bias logarithmic singularity, see Fig. M2a). Within the single QP tunneling scenario this is only possible if the depairing factor increases with $T$ at such a rate that it smears out the gap singularity in DoS and thus suppresses development of the singularity. We estimate that the corresponding $\Gamma$ at $T=T_c$ should be in the range 2-5 meV, somewhat smaller than deduced from ARPES \cite{Norman,Lee2007}.

Now we can substantiate our statement. In Fig. S8 we show the attempt of maintaining the same slope of $\ln dI/dV(V)$ curves for progressively decreasing $\Delta$. Since increase of both $\Gamma$ and $T$ leads to filling-in of the zero bias dip, i.e., to decrease of the slope, the slope can be maintained only if $\Gamma$ and $T$ move in the opposite directions. It is impossible to maintain the slope under the physical requirements that $\Gamma$ increases and $\Delta$ decrease simultaneously with increasing $T$ because all of the three parameters work in the same direction - decrease the slope. The inset in Fig. S7 shows the calculated zero bias resistance vs. $1/T$ for incoherent, non-directional, $d-$wave single QP tunneling and different $\Gamma$. It is seen that a prominent excess resistance appears even in this extreme $d-$wave case and even for very large $\Gamma$. The dashed line shows that the excess resistance grows approximately as $T^{-2}$ in this case. The abrupt appearance of the excess QP resistance is a mere reflection of the abrupt opening of the superconducting gap at $T_c$. Therefore, the observed $T-$independent slope and the absence of excess resistance, observed for UD mesas, are inconsistent with the single QP tunneling and points towards the doping-dependent change in the interlayer transport mechanism.

\begin{figure}
\includegraphics[width=3.0in]{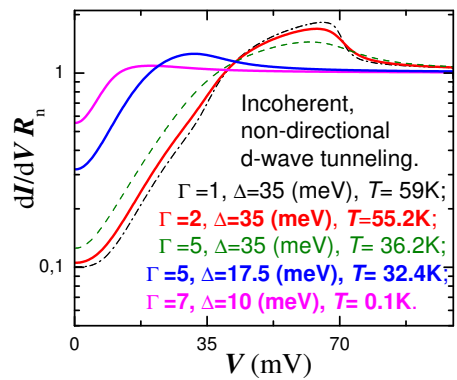}
\caption{\label{Fig1} (Color online). An attempt to maintain the slope of single QP characteristics at different $T$. It is seen that the slope can be maintained only under the un-physical requirement that $\Gamma$ increases and $\Delta$ decreases with decreasing $T$.}
\end{figure}

\subsection{Analysis of self-heating and non-equilibrium phenomena}

Self-heating can distort IVC's of Josephson junctions at large bias. The temperature rise due to self-heating is given by a simple expression
\begin{equation*}\label{Heat}
\Delta T = P R_{th}(T),
\end{equation*}
where $P=IV$ is the dissipated power and $R_{th}$ is the effective thermal resistance of the junctions, which is $T-$dependent and, therefore, bias dependent \cite{Heat}. The influence of self-heating on IVC's of our mesa structures was thoroughly studied in Refs.\cite{JAP,KrPhysC,Heat,HeatCom} (see also a discussion of scaling of IVC's in the inset of Fig. 1 and Fig. 3 from Ref.\cite{KrTemp}).

Despite relative simplicity of the phenomenon, discussion of self-heating in intrinsic tunneling spectroscopy has caused a considerable confusion, a large part of which has been caused by a series of publications by V.Zavaritsky \cite{Zavar}, in which he "explained" the non-linearity of intrinsic tunneling characteristics by assuming that there is no intrinsic tunneling. The irrelevance of this model to our subject was discussed in Ref.\cite{HeatCom}.

A certain confusion might be also caused by a large spread in $R_{th}$, reported by different groups \cite{Gough,AYreply,Heat,WangH,Lee}. For the sake of clarity it should be emphasized that those measurements were made on samples of different geometries. It is clear that $R_{th}$ depends strongly on the geometry \cite{JAP,Heat}, e.g., $R_{th}$ can be much larger in suspended junctions with poor thermal link to the substrate \cite{WangH} than in the case when both top and bottom surfaces of the junctions are well thermally anchored to the heat bath \cite{Lee}.
For mesa structures similar to those used in this study (a few $\mu m$ in-plane size, containing $N\simeq 10$ IJJ), there is a consensus that $R_{th}(4.2K) \sim 30-70 K/mW$ (depending on bias) \cite{AYreply,Heat} and $R_{th}(90K)\sim 5-10 K/mW$ \cite{Heat}. Larger values $R_{th} > 100K/mW$ claimed by some authors \cite{Gough} are unrealistic for our mesas because they can withstand dissipated powers in excess of $10 mW$ without being melted.

Yet, we note that talking about a typical value of $R_{th}$ is equally senseless as talking about a typical value of a contact (Maxwell) electrical resistance: both depend on the geometry. Therefore, reduction of mesa sizes provides a simple way for reduction of self-heating \cite{JAP}. Consequently, variation of $dI/dV$ characteristics with the junction size and geometry provides an unambiguous way of discriminating artifacts of self-heating from the spectroscopic features \cite{Heat}.

\begin{figure}
\includegraphics[width=3.8in]{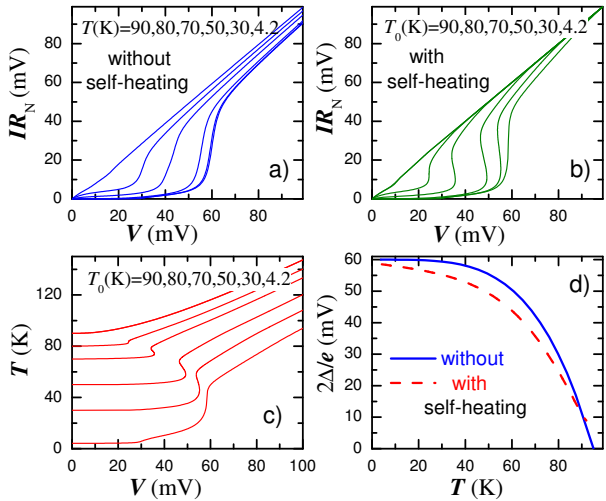}
\caption{\label{FigS7} (Color online). Simulated distortion of SIS tunneling characteristics by strong self-heating.
a) Undistorted IVC's at different $T$. Simulations were made for typical parameters of our mesas, $T-$dependent thermal conductivity of Bi-2212, and for coherent, directional, $d-$wave tunneling. b) distorted IVC's at the same base temperatures; c) The mesa temperature as a function of bias. d) Temperature dependence of the genuine superconducting gap (solid line) and the "measured" gap obtained from distorted IVC's (dashed curve). Note that even strong self-heating ($T$ reaches $T_c/2$ at $V_g$ at 4.2K) does not cause considerable distortion of the measured gap. Data from Ref.{\cite{KrPhysC}}}
\end{figure}

How self-heating {\it can} distort the IVC's of Josephson junctions is obvious: since self-heating rises the effective $T$ it may affect the IVC only via $T-$dependent parameters. There are three such parameters: the quasiparticle resistance, the superconducting switching \cite{Kras_TA,Collapse} current and the superconducting gap. They will affect the IVC's of Bi-2212 mesas, containing several stacked Josephson junctions, in the following manner:

The consecutive increase of $T$ upon sequential switching of IJJ's from the superconducting to the resistive state will distort the periodicity of quasiparticle branches. Each consecutive QP branch will have a smaller QP resistance (smaller $V$ at given $I$) and smaller switching current, see the discussion in Refs. \cite{HeatCom,KrTemp}. This type of distortion of the QP branches becomes clearly visible (at base $T=4.2K$) when $\Delta T \gtrsim 20K$ \cite{JAP,HeatCom}.

The $T-$dependence of $\Delta$ may lead to appearance of back-bending of the IVC at the sum-gap knee.
Fig. S9 reproduces the results of numerical simulation of such the distortion, made specifically for the case of Bi-2212 mesa with the corresponding $T-$dependent parameters (see Ref.\cite{KrPhysC} for details). Fig. S9 a) shows a set of simulated undistorted IVC's at different $T$ for coherent, directional, $d-$wave tunneling with a $\Delta(T)$, shown by the solid line in Fig. S9 d). Panels b) and c) show the distorted IVC's and the actual junction temperature, respectively. The dashed line in panel d) represents the "measured" gap obtained from the peak in distorted $dI/dV$ characteristics. Remarkably, the deviation from the true $\Delta(T)$ is marginal, despite large self-heating, $\Delta T \simeq T_c/2$ at $4.2 K$!

The robustness of the measured gap with respect to self-heating is caused by a flat $T-$dependence of the superconducting gap at $T<T_c/2$. If self-heating reaches $T_c$ at the sum-gap knee, an acute back-bending appears in the IVC's, however even this does not cause principle changes in the "measured" $\Delta(T)$. In experiments on large mesas \cite{Gough} or suspended structures \cite{WangH}, in which acute self-heating was reported, no clear Ohmic tunneling resistance could be observed at high bias, in stark contrast to our mesas, see Fig. M1a, and Refs.\cite{KrTemp,KrMag,Doping}.

Simulations as in Fig. S9 clearly show that irrespective of self-heating the "measured" gap vanishes at $T=T_c$, i.e., simultaneously with the true gap. Thus trivial self-heating simply can not "hide" the qualitative $\Delta(T)$ dependence. For example, there is no way in which one can get the vanishing "measured" (self-heating affected) gap if the true gap is $T-$independent. Therefore, the discrepancy in the measured strong $T-$dependence of the gap at $T_c$ in intrinsic tunneling (and recent ARPES \cite{Lee2007}), or complete $T-$independence in STM measurements \cite{Renner} can not be attributed to self-heating.

To quantitatively analyze the significance of self-heating in our mesas, we provide values of the dissipation power $P=IV$ for the studied mesas: For the near-OP Bi(Y)-2212 mesa at $T=4.9K$ dissipation at the sum-gap peak is $0.82 mW$. From the previous analysis of the size dependence of $V_g$ for similar mesas \cite{Heat} it was observed that $V_g$ becomes size-independent, and therefore not affected by self-heating, for small mesas with $P(V_g) < 1mW $. All the mesas studied in this manuscript fall into this category. All of them exhibit perfect periodicity of quasiparticle branches \cite{KrTemp} and none of them exhibit back-bending at any $T$, which according to Fig. S9, implies that self-heating at $V_g$ is less than $T_c/2$ at $T=4.2K$. This is consistent with previous in-situ measurements \cite{Heat}, $R_{th}(T=4.2K,V=V_g) \simeq 30-40K/mW$.

The rest of the data presented in the manuscript is not affected by self-heating for the following reasons:

The effective $T_{eff}$ in Fig. M2b) was obtained at $V/N = 30meV$ at which the dissipation power was only $P \simeq 3.8 \mu W$ at $T=5.5K$ (for comparison, $P(V_g)=0.21 mW$ for the same curve). Therefore, self-heating here is negligible (sub-Kelvin). Besides, as argued in the manuscript, the slope of $\ln dI/dV(V)$ in Fig. M2 is apparently $V$ and $P$ independent, therefore the same data could be obtained at much smaller or larger $P$. The data in Figs. M3 b-d) and M4 is free from self-heating because it was obtained at zero bias.

Finally, it is important to emphasize that the concept of heat diffusion is inapplicable for small Bi-2212 mesas containing only few atomic layers. The phonon transport in this case is ballistic \cite{Heat,Uher} and the energy flow from the mesa is determined not by collisions between the tunneled non-equilibrium QP with thermal phonons, but by spontaneous emission of a phonon upon relaxation of the non-equilibrium QP \cite{Cascade}. This process is not hindered at $T=0$. Therefore, the effective $R_{th}$ (and self-heating) can be much smaller because it is not limited by poor thermal conductivity at $T=0$, but is determined by the fast, almost $T-$independent, non-equilibrium QP relaxation time. The concept of self-heating becomes adequate only in the bulk of the Bi-2212 crystal, where the dissipation power density and the temperature rise are much smaller due to the much larger area of the crystal. For more details see the discussion in Ref. \cite{Heat}. The non-equilibrium energy transfer channel is specific for atomic scale intrinsic Josephson junctions made of perfect single crystals. It can explain a remarkably low self-heating at very high bias \cite{Cascade}.

\subsection{Thermal activation analysis}

\begin{figure}
\includegraphics[width=3.0in]{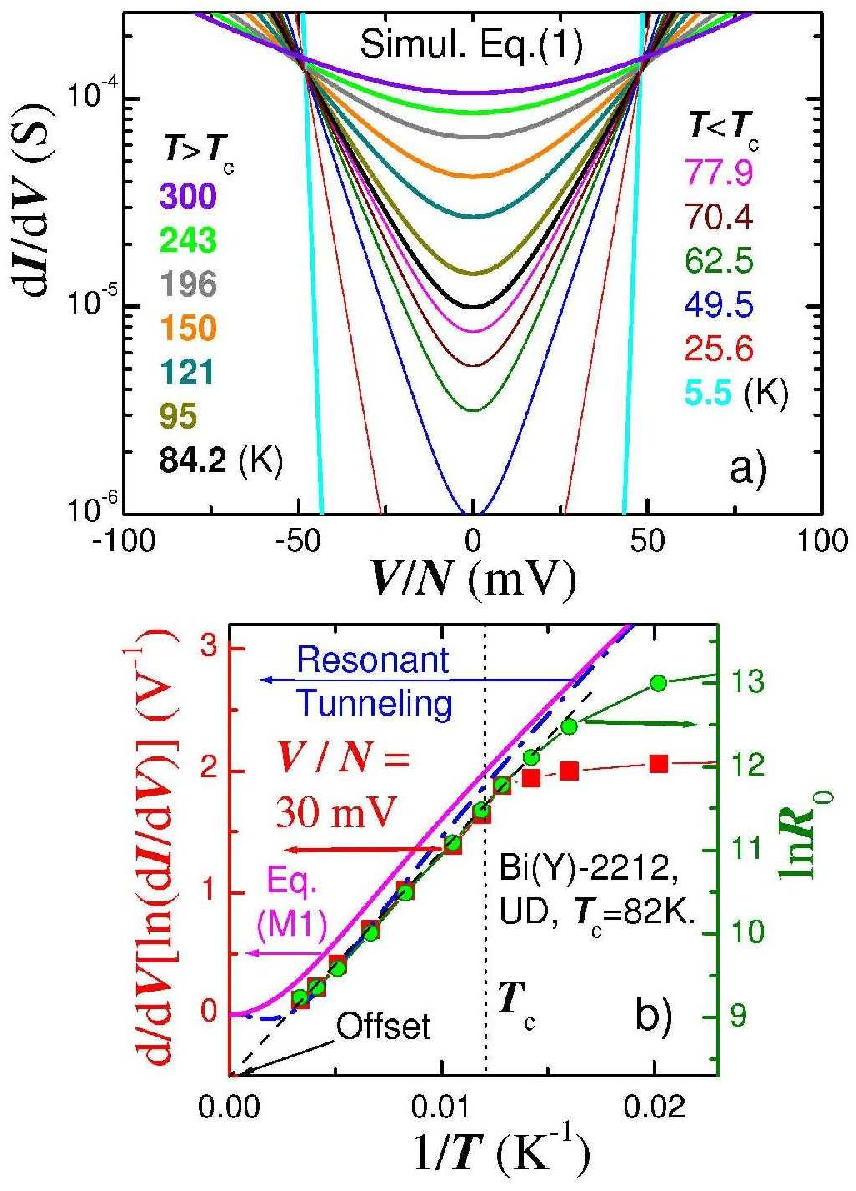}
\label{FigS6}
\caption{(Color online). a) Simulated $dI/dV$ characteristics according to Eq.(M1) for the same $T$ as in Fig. M2 a). It is seen that at $T>T_c$ they provide a good fit to experimental data, including the crossing point. b) The slope of experimental characteristics from Fig. M2 a) at $V/N=30mV$ (red squares, left axis) and the logarithm of zero bias resistance (green circles, right axis) vs. $1/T$. It is seen that both follow the TA behavior at $T>T_c$ and simultaneously deviate downwards in the superconducting state, indicating that appearance of the negative excess resistance and the crossover to quantum $c-$axis transport are correlated. The dashed-dotted line represents the slope of resonant tunneling characteristics (shown for comparison).}
\end{figure}

The TA current is given by a simple expression:
\begin{equation}
I_{TA} \propto n(T) \exp\left[-\frac{U_{TA}}{k_B T}\right]\cosh\left[\frac{eV}{2k_B T}\right], \label{ITA}
\end{equation}
from which follows Eq.(M1).

Fig. S10 a) shows the $dI/dV(V)$ characteristics in a semi-log scale calculated from Eq.(M1) for the same $T$ and scale as in Fig. M2 a), in order to facilitate the direct comparison. Simulations were made for constant $\sigma_N$ and $U_{TA} = 24 meV$, which were adjusted to obtain similar $R_0(T)$ values as in the experimental data from Fig. M2 a) at $T>T_c$; and for $n(T) \propto T$. Considerable grows of mobile carrier concentration $n(T)$ with $T$ in UD crystals was reported in Hall effect measurements \cite{Hall}. The assumption of linear $n(T)$ is supported by observation that for strongly UD mesas it is $R_0$ (circles in Fig. S10 b), rather than $R_0/T$, that is more accurately described by the Arrhenius law (dashed line).

From comparison of Figs. M2a) and S10a) it is seen that Eq.(M1) does provide a good fit to experimental data at $T>T_c$, including the crossing point. Simultaneously a dramatic discrepancy is seen below $T_c$. This indicates that the slopes of $dI/dV(V)$ curves in the superconducting state are not determined by the real $T$, but by some constant effective temperature, as shown in Fig. M2 b).

Fig. S10 b) shows the slope of experimental curves from Fig. M2a) at $V/N = 30 mV$ (red squares) and the logarithm of zero bias resistance (green circles) as a function of $1/T$. Clear linear dependence of both at $T>T_c$ indicates that the IVC's in the normal state have thermal activation nature both at zero and finite bias, consistent with Eq.(M1). According to Eq.(M1) the effective temperature can be explicitly obtained from the slope $d/dV[\ln(dI/dV)]=$ $(e/2k_BT) \tanh (eV/2k_BT)$ $\simeq e/2k_BT$. Here there are two slight obstacles: First, the $tanh$ term deviated from unity at high $T$ leading to eventual saturation of the slope at $1/T\rightarrow 0$. The exact $T-$depensence according to Eq.(M1) is shown by the solid (magenta) line in Fig. S10 b). We emphasize that there are no fitting parameters in this curve. It is seen that Eq.(M1) reproduces quite well $T-$dependence of the slopes, except for a small offset. This additional offset is caused by the fact that the slope of experimental curves becomes negative at large $T$ \cite{Doping}. The simple TA expression, Eq.(M1), does not reproduce this negative slope. However, more advanced TA simulations do reproduce the negative slope and, in fact, all the details of experimental $dI/dV$ characteristics at $T>T_c$ \cite{Silva}.

It should be noted that TA-like behavior is quite universal for many process. Except for pure thermal activation over the finite barrier without the gap in the electronic DoS \cite{Silva}, it may appear in pure tunneling characteristics in the presence of the gap in electronic DoS, due to TA-like behavior of the Fermi factor (even though this would require a very specific correlation between $T-$dependent factors mentioned in sec.I); as a result of inelastic tunneling via the impurity \cite{Impurity}, or elastic tunneling via a resonant state \cite{Abrikosov} in the tunnel barrier. Abrikosov has shown that the later scenario can quantitatively reproduce the interlayer characteristics of HTSC in the normal state \cite{Abrikosov}. The blue dash-dotted line in Fig. S10 b) represents the slope $d/dV[ln(dI/dV)]$ in the case of resonant tunneling with the appropriate energy of the resonant state. It follows the simple linear TA behavior at $k_B T$ lower than the resonant energy and saturates at higher $T$. Comparison with the resonant tunneling calculation shows that the saturation of the slope at high $T$ results in appearance of a negative offset in the linear $d/dV[ln(dI/dV)]$ vs $1/T$ dependence at $1/T\rightarrow 0$. Below the saturation temperature, the actual temperature can be easily extracted from the $d/dV[ln(dI/dV)]$ slopes, compensated by the offset at $1/T \rightarrow 0$, as indicated in Fig. S10 b) :
\begin{equation*}\label{TeffOff}
T_{eff} = \frac{e}{2k_B \left[d/dV[\ln(dI/dV)]-Offset(T^{-1}\rightarrow 0)\right]}.
\end{equation*}
This expression was used for obtaining the effective $T$ shown in Fig. M2b).

From Fig. S10 b) it is seen that at $T<T_c$ both $\ln R_0$ and the slope of experimental curves simultaneously deviate downwards with respect to the linear TA behavior. Therefore, both are consequences of the same phenomenon (which we attributed to doping-induced change in the $c-$axis transport) only at zero and finite bias, respectively.

\end{document}